\documentclass{article} 
\usepackage[preprint]{colm2025_conference}

\usepackage{microtype}
\usepackage{hyperref}
\usepackage{url}
\usepackage{booktabs}
\usepackage{graphicx} 
\usepackage{wrapfig}
\usepackage{lineno}
\usepackage{pifont}
\usepackage{lipsum}
\usepackage{amsmath}
\usepackage{xcolor}
\usepackage{multicol} 
\usepackage{subcaption}
\usepackage{array}
\usepackage{multirow}
\usepackage{enumitem}
\usepackage{pifont}
\usepackage{hyperref}
\usepackage{tikz}
\usepackage{tcolorbox}

\definecolor{darkblue}{rgb}{0, 0, 0.5}
\hypersetup{colorlinks=true, citecolor=darkblue, linkcolor=darkblue, urlcolor=darkblue}

\definecolor{burntorange}{rgb}{0.8, 0.33, 0.0}

\definecolor{teal}{rgb}{0.0, 0.6, 0.6}

\title{
Self-Resource Allocation in Multi-Agent LLM Systems}

\author{Alfonso Amayuelas$^1$, Jingbo Yang$^1$, Saaket Agashe$^2$, \\
\textbf{Ashwin Nagarajan$^2$, Antonis Antoniades$^1$, Xin Eric Wang$^2$, William Wang$^1$}\\ 
$^1$University of California, Santa Barbara\\
$^2$University of California, Santa Cruz\\
\texttt{\{amayuelas,jingbo,antonis\}@ucsb.edu, \{saagashe,asnagara\}@ucsc.edu }  \\
\texttt{xwang366@ucsc.edu}, \texttt{william@cs.ucsb.edu} \\
}

\begin{document}

\ifcolmsubmission
\linenumbers
\fi

\maketitle

\begin{abstract}
\vspace{-10pt}
With the development of LLMs as agents, there is a growing interest in connecting multiple agents into multi-agent systems to solve tasks concurrently, focusing on their role in task assignment and coordination. This paper explores how LLMs can effectively allocate computational tasks among multiple agents, considering factors such as cost, efficiency, and performance. In this work, we address key questions, including the effectiveness of LLMs as orchestrators and planners, comparing their effectiveness in task assignment and coordination. Our experiments demonstrate that LLMs can achieve high validity and accuracy in resource allocation tasks. We find that the planner method outperforms the orchestrator method in handling concurrent actions, resulting in improved efficiency and better utilization of agents. Additionally, we show that providing explicit information about worker capabilities enhances the allocation strategies of planners, particularly when dealing with suboptimal workers.
\vspace{-10pt}
\end{abstract}


\section{Introduction}

Large Language Models (LLMs) have become popular for various tasks beyond just generating text. They can act as agents that interact with their environment \citep{survey_agents} and use tools effectively. This has led to the development of more versatile systems that can operate a computer like a human \citep{agent_s, wu2024copilot}, serve as research assistants \citep{agent_research_assistant}, or even automate tasks in a lab \citep{chemcrow}.

As agents are used for more tasks and in more scenarios, they need to interact with other agents. This can happen by design in Multi-Agent Systems (MAS) or through more spontaneous interactions. As a result, frameworks like AutoGen \citep{autogen} and Camel-AI \citep{camel_ai_framework} have been developed to leverage the capabilities of MAS. The creation of these multi-agent systems opens up new possibilities and challenges to explore \citep{han2024llm, guo2024large, llm_coordination_benchmark}, such as their organizational structures, shared memory, communication efficiency, and coordination capabilities.

In particular, we focus on how these agents coordinate to assign tasks and achieve their goals. Inspired by Marvin Minsky's idea that intelligence emerges from computational modules working together to accomplish goals that none could achieve alone \cite{minsky1988society}, we aim to analyze how LLMs allocate tasks to agents. Our goal is to optimize the allocation of resources and tasks by LLMs themselves. Thus, our underlying research question is: \textbf{How does a network of LLM-based agents optimize their task allocation?}

\begin{figure}[t!]
    \centering
    \includegraphics[width=\linewidth]{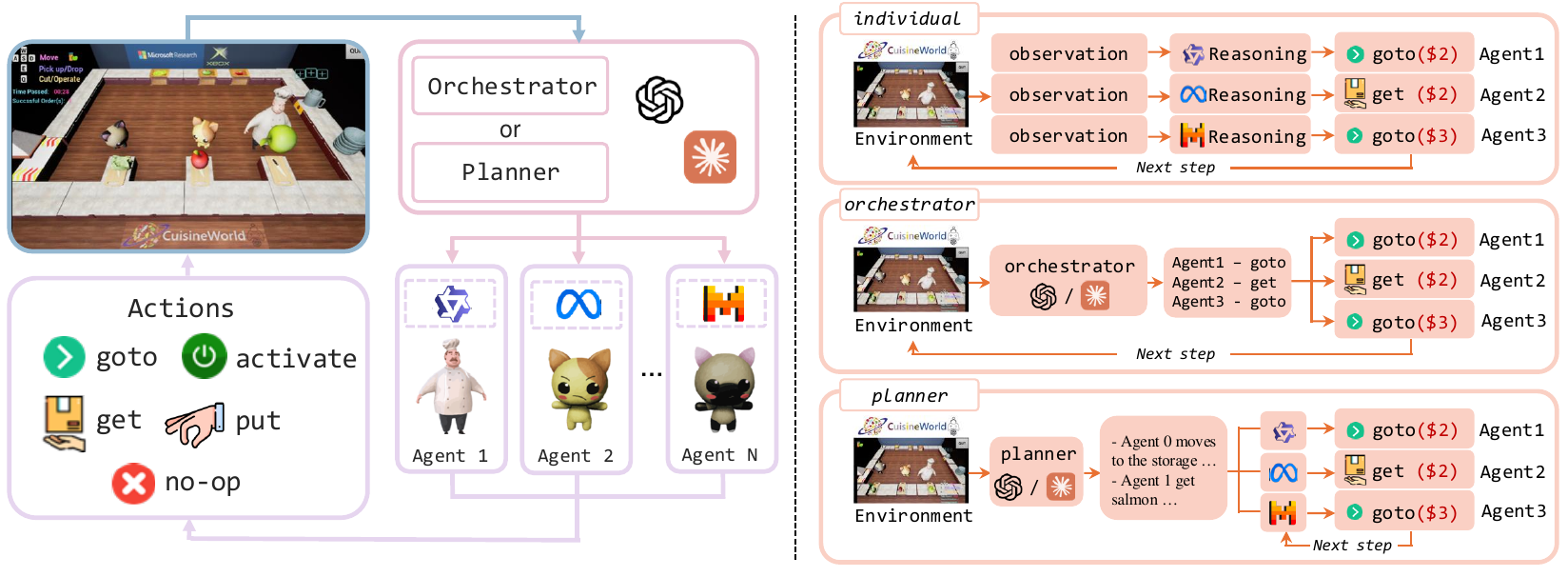}
    \caption{MultiAgents Systems on CuisineWorld \citep{mindagent} over 3 methods analyzed: (1) \textbf{Individual}: Decentralized, (2) \textbf{Orchestrator}: Centralized, (3) \textbf{Planner}: A balanced combination. It generates a plan every few steps which is then given to worker agents to generate their actions.}
    \label{fig:general_image}
    \vspace{-10pt}
\end{figure}

We conduct a series of experiments to understand how LLMs allocate tasks to agents within multi-agent systems. First, we evaluate how one LLM can generate the correct task allocation, assigning actions to each agent for a problem with a known solution provided by the Hungarian Algorithm.  Second, we examine two methods illustrated in Figure \ref{fig:general_image}—Orchestrator and Planner—using the CuisineWorld benchmark  \citep{mindagent}, which is detailed in Section \ref{section:experiment2}. The Orchestrator uses one LLM to generate all the actions to be executed, while the Planner creates a plan that is then given to executor LLM agents, who generate their actions independently. The plan is only re-evaluated when a relevant event occurs. Lastly, we assess how the Planner can allocate tasks based on the agents' abilities.


From the experiments, we concluded interesting findings: First, LLMs achieve higher validity and accuracy in resource allocation tasks as their parameter size increases, but at a significant computational and monetary cost. Second, the planner method outperforms the orchestrator method in handling concurrent actions, resulting in improved efficiency in multi-agent task execution. The planner method achieves better utilization of agents, with fewer idle actions. Finally, we see that LLMs are sensitive to worker capabilities and struggle to infer them dynamically. Providing explicit information about worker capabilities improves planner's allocation strategies, especially with suboptimal workers.

We highlight the following key contributions from this paper:

\begin{itemize}[leftmargin=*]
    \item An evaluation of LLMs as effective orchestrators for optimizing task allocation in multi-agent systems by comparing their task allocations against the Hungarian Algorithm optimal solution.
    \item We demonstrate the differences between the orchestrator and the planner methods when handling concurrent actions, highlighting the efficiency gains achieved by the planner.
    \item An analysis of how the Planner allocates tasks based on agents' abilities within the system, demonstrating that providing explicit information about worker capabilities enhances allocation strategies, particularly when dealing with suboptimal workers.
\end{itemize}

As multi-agent systems evolve, optimizing task allocation will be critical for enhancing efficiency and performance across applications. This work addresses current resource management challenges in LLM-based frameworks and paves the way for advancements in autonomous and collaborative AI systems.

\section{Related Work}

\paragraph{Agent Interaction and Collaboration} As more agents are deployed in real-world settings, studying their interactions becomes increasingly relevant. These agents need to communicate not only with each other but also with humans \citep{jiang2025experimentalexplorationinvestigatingcooperative, liu2023llm}. Agents can choose to cooperate, compete, or do a mix of both \citep{tran2025multi}. Since these models use natural language for communication, recent studies have explored the concept of "Theory of Mind," which involves understanding and attributing \textit{mental} states to oneself and others \citep{strachan2024testing, street2024llm}. Some works apply Game Theory to analyze these interactions \citep{hua2024gametheoreticllmagentworkflow}. Several studies measure the coordination abilities of LLMs, such as the LLM-Coordination Benchmark \citep{llm_coordination_benchmark} and MindAgent \citep{mindagent}, Gamma($\gamma$)-Bench \citep{gamma-bench}. Other works focus on improving the abilities of LLMs to interact and coordinate \citep{li2023theory, zhang2024towards, zhang2023exploring, cross2024hypothetical, zhang2023building, guo2024embodied}.

\paragraph{LLM-based Multi-Agent Systems} Extending single-agent systems to multi-agent systems has led to increased computational efficiency during inference, with shorter execution times. Multi-agent systems offer several benefits, including modularity, specialization, collaborative learning, and improved decision-making. These systems provide better scalability and flexibility, making them more effective at solving problems that a single model might struggle with. The development of multi-agent systems has led to the creation of various frameworks focused on MAS, such as AutoGen \citep{autogen}, Camel-AI \citep{camel_ai_framework}, or MetaGPT \citep{metagpt}. These frameworks have enabled the creation of systems with better scalability and flexibility, enhancing problem-solving capabilities. For example, the Chain of Agents \citep{zhang2024chain} can process long contexts effectively, FilmAgent \citep{filmagent} generate coherent long-sequence videos , Chemrow \citep{chemcrow} support new chemistry research, 
and Smart-LLM to control multiple robots simultaneously \citep{kannan2024smart}.

\paragraph{Learning to optimize resources} When developing multi-agent systems, a crucial question arises: What is the optimal architecture? This is explored in \citet{zhuge2024gptswarm, liu2023dynamic}. Once the architecture is defined, how do these systems utilize it effectively? First, selecting the right model for specific queries can optimize performance by ensuring high-quality answers, as explored in RouteLLM \citep{route-llm} Hybrid-LLM \citep{hybrid-llm}, and the corresponding benchmark RouterEval \citep{huang2025routerevalcomprehensivebenchmarkrouting}. In Mindstorms \citep{mindstorms}, the authors theorize how LLMs can lead to systems that self-manage resources based on a monetary system, optimizing resources and maximizing rewards to create an "\textit{Economy of Minds}" (EOM). Models can also learn to optimize their communication, as demonstrated in Agora \citep{marro2024scalable}. Due to the need for communication and coordination, Multi-Agent Reinforcement Learning (MARL) becomes more complex than simply scaling up single-agent RL \citep{sun2024llm}. Self-organized MASs can reduce the burden on developers while achieving better results, such as in code generation \citep{ishibashi2024selforganizedagentsllmmultiagent} or leveraging scale in multi-agent reward-based learning \citep{fama, 2024arXiv241006101M, tekin2025multiagentreinforcementlearningfocal}.

\section{Problem Definition}

In real-world settings, resource allocation in multi-agent systems can often be seen as a multi-agent resource allocation problem. This involves distributing computational tasks among multiple model instances to optimize criteria such as cost, efficiency, and performance. This challenge becomes especially relevant as AI systems gain autonomy and need to make resource allocation decisions with incomplete information. We formalize the problem to provide a foundation for analyzing how LLMs can perform self-resource allocation under multiple constraints and objectives.

In a multi-agent system with $N$ agents and $P$ different tasks, each task can be decomposed into a sequence of $M_p$ subtasks. Each agent incurs different costs for different tasks due to variations in capability. The objective is to maximize overall utility while respecting time and assignment constraints.

\paragraph{Agents} Let $\mathcal{A} = \{a_1, a_2, \ldots, a_n\}$ represent a set of agents, where agent is characterized by
\begin{enumerate}[nosep]
    \item Operational Cost ($c_i$): It represents the computational expense.
    \item Capability Measure ($\phi_i$): It reflects the model proficiency in performing tasks.
\end{enumerate}

\paragraph{Tasks} $\mathcal{T} = \{t_1, t_2, \ldots, t_p\}$ denote a set of tasks to be allocated. A task $t_j \in \mathcal{T}$ is defined by:

\begin{enumerate}[nosep]
    \item Difficulty level ($d_j$): It indicates the task complexity
    \item Sub-tasks ($m_j$): It refers to the subtask that a main task can be decomposed into.
    \item Workload requirement ($w_j$): Representing computational demand
    \item Reward ($r_j$): A potential reward, which may not be available in all scenarios.
\end{enumerate}

\paragraph{Utility Function} Combine utility function for a subtask
\begin{equation}
   \begin{cases}
    q_{pim}-c_{pim}, & \text{if agent } i \text{ can execute subtask } m \text{ for task } p \\
    -\infty, & \text{otherwise}
\end{cases} 
\end{equation}
\noindent where $q_{pim}$ and $c_{pim}$: Quality and cost, respectively, for allocating agent $i$ to work on subtask $m$ for task $p$.

The assignment of subtask $m$ to agent $i$:
    \begin{equation}
    v_{pim}= 
    \begin{cases}
    1, & \text{agent } i \text{ is assigned to subtask } m \text{ for task } p \\
    0, & \text{otherwise}
     \end{cases}
    \end{equation}

\paragraph{Optimization Problem}  The goal is to maximize the utility under a time constraint $T_{max}$:

\begin{equation}
\arg\max_{v} \sum_{p=1}^{P} \sum_{i=1}^{N} \sum_{m=1}^{M_p} u_{pim} v_{pim}
\end{equation}

Subject to:
\begin{align}
\sum_{p} \sum_{i} \sum_{m} \tau_{pim} v_{pim} &\leq T_{max} \\
\sum_{i} v_{pim} &\leq 1 \quad \forall m \in \mathcal{M}_p, \forall p \in \mathcal{P} \\
v_{pim} &\in \{0, 1\} \quad \forall i \in \mathcal{N}, \forall m \in \mathcal{M}_p, \forall p \in \mathcal{P}
\end{align}

As noted by \citet{korsah2013comprehensive, mindagent}, this problem cannot be solved in polynomial time. In our work, we aim to tackle this problem with LLMs.

\paragraph{Assignment problem}

For the basic assignment problem in Experiment 1, we can simplify the notation. We set $N = P$, indicating that the number of agents equals the number of tasks. Each task consists of a single subtask, represented by $M_p = 1$. The cost matrix $A \in \mathbb{R}^{n \times n}$ represents the costs, with $c_{pi1} = A_{ij}$. We focus solely on minimizing costs, setting $q_{pi1} = 0$. Each assignment takes one unit of time, denoted by $\tau_{pi1} = 1$. Finally, we set $T_{max} = N$, ensuring that the time constraint allows each agent to be assigned exactly one task.

\begin{equation}
\arg\min_{v} \sum_{i=1}^{N} \sum_{j=1}^{N} A_{ij} v_{ij}
\end{equation}

$\text{s.t.} \quad \sum_{i=1}^{N} v_{ij} = 1 ;$ $ \forall j, ; \sum_{j=1}^{N} v_{ij} = 1 \forall i,  $ ensuring each agent is assigned to one task\\
$v_{ij} \in {0, 1} ; \forall i,j \in {1, \ldots, N}$ specifying the assignment variable is binary.

\begin{table}[h]
\centering
\small
\begin{tabular}{p{3cm}p{3cm}p{3cm}p{3cm}}
\toprule
& \textbf{Experiment 1} & \textbf{Experiment 2} & \textbf{Experiment 3} \\
\midrule
Objective & Minimize total cost & Maximize completed tasks & Maximize completed tasks \\
\midrule
Environment & Static assignment problem & Dynamic CuisineWorld & Dynamic CuisineWorld with varying agent capabilities \\
\midrule
Reward Structure & Explicit costs & Delayed rewards & Delayed rewards \\
\midrule
Agent Capabilities & Uniform & Uniform & Varied \\
\midrule
Decision Making & Centralized & Centralized (Orchestrator) vs. Semi-Decentralized (Planner) & Semi-Decentralized (Planner) with capability awareness \\
\midrule
Evaluation Metric & Accuracy and Validity rate & Task Completion Rate and Efficiency & Task Completion Rate and Efficiency \\
\bottomrule
\end{tabular}
\caption{Comparison of the three experimental scenarios}
\label{tab:experiment_comparison}
\end{table}

This formulation allows us to evaluate how effectively LLM orchestrators can allocate resources when costs are explicitly defined, as in the Hungarian algorithm comparison.

\section{Experimental Framework}

Our problem formulation includes three experimental scenarios of increasing complexity:

\subsection{Experiment 1: Basic Resource Allocation}

\paragraph{Problem Statement} Orchestrator agent is widely adopted in multi-agent systems. We start with evaluating how effectively LLM orchestrator can allocate resources in the most simple setting where both costs and rewards are explicitly defined. 
\begin{wrapfigure}{r}{0.5\linewidth}   
  \centering
  \includegraphics[width=\linewidth]{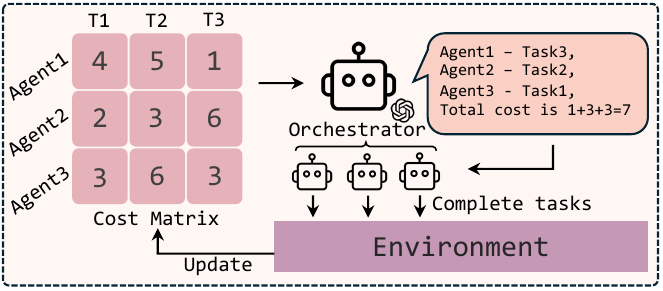}
  \caption{A multi-agent system with an LLM-based orchestrator. At each step, the orchestrator optimally assigns multiple tasks to agents and minimizes the total cost. After agents complete their tasks, the environment updates with new tasks and a new cost matrix, continuing the process iteratively.}
  \label{fig:exp1}
  \vspace{-10pt} 
\end{wrapfigure}
Consider a multi-agent system with a long-horizon task that can be broken down into a sequence of fundamental sub-tasks or actions executable by agents. In this system, the LLM-based orchestrator assigns tasks to agents based on their associated costs. Each agent incurs different costs for different tasks due to variations in capability or access to resources. Once an assignment is made and agents complete their tasks, the environment updates accordingly, leading to a new set of tasks for the next turn, as is shown in Figure~\ref{fig:exp1}. Here, we focus on evaluating the orchestrator LLM in a single turn, aligning with the standard assignment problem.

The objective of the assignment problem is to determine the optimal assignment of $n$ agents to $n$ tasks. Each instance of the problem is defined by a cost matrix $A \in \mathbb{R}^{n \times n}$, where each entry represents the cost associated with assigning a particular agent to a task.
As a fundamental combinatorial optimization problem, the assignment problem can be solved in polynomial time using the Hungarian algorithm, which has a complexity of $O(n^3)$, where $n$ is the number of tasks and agents. Here, we aim to evaluate whether LLMs can effectively minimize the assignment cost out of the box.

\paragraph{Experiment Details} 
To evaluate the LLM performance on assignment problems, we first created an evaluation set by generating a multiple cost matrix with random numbers filled in. After that, we utilize the Hungarian algorithm to get the ground truth for each answer. All LLM orchestrators are evaluated using greedy decoding with a maximum sequence length of 2048.

\paragraph{Evaluation}
We evaluate the performance of self-resource allocation across LLMs of varying model sizes. Specifically, we consider GPT-4o-mini(\textasciitilde8B), Mistral-Small-3.1(24B), Qwen2.5-32B-Instruct, Llama-3.1-70B-Instruct, GPT-4o(\textasciitilde200B), and Llama3.1-405B-Instruct-FP8. 
Two evaluation metrics, accuracy and validity rate, are used. \ding{182} \textbf{Accuracy} measures how many of the LLM-generated assignment solutions are optimal. A solution is considered as correct only when it is exactly the same as the ground truth optimal assignment given by the Hungarian algorithm. Here we use the LLM judge based on GPT-4o to examine each model response.
\ding{183} \textbf{Validity rate} tests how many of the LLM-generated assignment solutions are valid, which can be not optimal. Invalid assignments include assigning one agent to more than one task, leaving some tasks not assigned with agents, and making up an incorrect lower cost of the agent.

\subsection{Experiment 2: Concurrent Allocation}
\label{section:experiment2}

\paragraph{Problem Statement} In this experiment, we introduce additional complexity by incorporating delayed rewards, as outlined in Table \ref{tab:experiment_comparison}. The reward is only provided after the completion of a number of actions. Rewards  In real-world scenarios, agents often need to execute actions without immediate knowledge of whether those actions will ultimately.

\begin{wrapfigure}{r}{0.5\linewidth}   
  \centering
  \includegraphics[width=\linewidth]{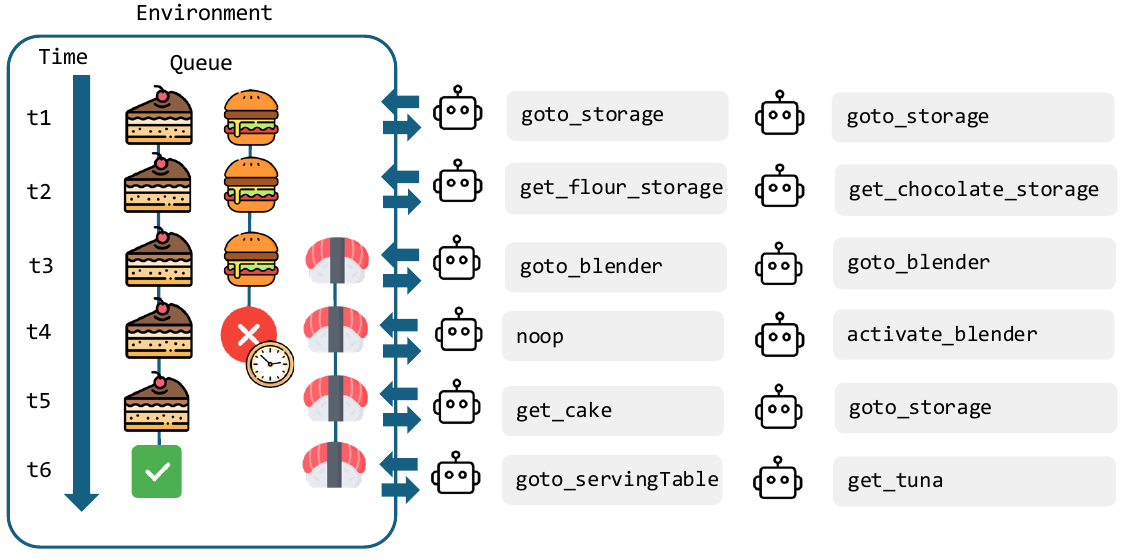}
  \caption{In CuisineWorld, the agents are presented with a set of dishes to complete in certain. Agents execute tasks until dishes are completed or the order expires.}
  \label{fig:exp2_cuisineworld}
    \vspace{-20pt} 
\end{wrapfigure}

We use CuisineWorld \citep{mindagent} as the benchmark, inspired by Overcooked! \citep{carroll2019utility}. Agents must complete dish orders by collecting and cooking ingredients, then delivering them within a time frame. The goal is to maximize completed tasks. The benchmark features 10 locations, 27 ingredient types, and 33 dishes, categorized by difficulty and cooking tools required, resulting in 12 game levels. Figure \ref{fig:exp2_cuisineworld} represents this game setting. We refer to Appendix \ref{app:cuisineworld} for more information.

As originally described, the agent decision process can be formulated as a \textit{Markov Decision Process} ($\mathcal{S, A, T, M, G}$) with state space $\mathcal{S}$, action space $\mathcal{A}$, (effectively indicating all the possible schedules that can be made at a single time step), transition dynamics $\mathcal{T}$ , reward function $\mathcal{R}$ and task instruction space $\mathcal{G}$. 

\noindent \textbullet State Space $\mathcal{S}$: The environment consists of two main entities: locations and agents. Locations can be storage areas, serving tables, or cooking tools like pans and blenders. Each location's description includes the items it contains and whether it is currently occupied. Agents are described by their current location, the items they are holding, and whether they are using a tool.

\noindent \textbullet Actions: Actions in CuisineWorld involve dispatching commands to agents, such as moving to a location, obtaining or placing items, activating tools, or performing no operation. The list of actions include \texttt{goto} (agent, location), \texttt{get} (agent, location, item), \texttt{put} (agent, location), \texttt{activate} (agent, location) and the idle action.

\noindent \textbullet Tasks: The tasks involve completing dish orders, which range from simple to complex recipes. New tasks are introduced at regular intervals, and each task has a limited time frame within which it must be completed, otherwise it fails.

\paragraph{Experiment Details} The goal of this experiment is to evaluate the effectiveness of planning in a multi-agent LLM system compared to the use of an orchestrator. We follow the definition of an agent provided in React \citep{react} and other literature \citep{xi2025rise}, where agents are defined as: \textit{environment} $\rightarrow$ \textit{reasoning} $\rightarrow$ \textit{action}. We compare three methods, represented in Figure \ref{fig:general_image} from $n=1, ..., 6$ agents:
\begin{enumerate}
    \item \emph{Individual}: Each agent is controlled by a different LLM, and each agent generates its own action independently.
    \item \emph{Orchestrator}: One LLM controls all the agents in the game and generates actions for all of them.
    \item \emph{Planner}: Generates a general plan whenever a relevant event occurs in the game, such as when a dish is introduced, completed, or removed. This plan is then provided to the LLM agents, which produce their independent actions.
\end{enumerate}

\paragraph{Evaluation} We evaluate the performance of the previous methods in producing the correct allocation and maximizing the number of orders completed. We aim to study how individual, independent weak models (decentralized) can benefit from planning by a larger model. For this purpose, we use Llama 70B-Instruct, Qwen 32B-Instruct, and GPT4o-mini as executor models, while GPT-4o and Claude 3.7 Sonnet are used for the Orchestrator (centralized). The Planner method is then evaluated using a combination of these models, where GPT-4o and Claude 3.7 generate the plans, and open-source models generate the actions. To evaluate the results, we use the following metrics: \ding{182} \textbf{\#Completed Orders}: the number of orders the agents are able to complete; \ding{183} \textbf{Execution Cost}: the cost of LLM calls at current prices, detailed in Table \ref{tab:models_prices}, which closely represents their running cost; and \ding{184} \textbf{Efficiency}=$\frac{\#complted\_orders}{\$cost}$: the ratio of completed orders to cost, representing the work done per dollar spent.

\subsection{Experiment 3: Capability-Aware Allocation}


\paragraph{Problem Statement.} In the third experiment, we want to evaluate the capability of LLMs to allocate tasks based on the ability of the worker agents. We fix the Planner model and evaluate its ability to allocate plans to a heterogeneous mix of worker agents, each with a different backbone model. Moreover, each of these backbone models has varying levels of intelligence and parameter sizes -- allowing us to evaluate the task distribution of LLMs to Worker LLMs of varying intelligence capabilities. We use the same CuisineWorld environment \cite{mindagent} - with similar settings as Experiment 2 (Section \ref{section:experiment2}). 

\paragraph{Experiment Details}
We run two experimental settings to evaluate the Capability-Aware Allocation abilities of LLMs as Planners. In the first setting, \emph{On-the-fly Allocation}, the planner agent is not provided any information about its workers and needs to infer their capabilities dynamically during execution. In the second setting, \emph{Informed Allocation}, we want the Planner to have knowledge about the capabilities of the worker agents ($\phi_i$). We indicate the capabilities of underlying worker models. For this, we use the action success rates of the individual models on CuisineWorld as a proxy measure of their intelligence. 

\paragraph{Evaluation: } For the evaluation we use the Claude-3.7-sonnet model as the planner, as it shows to have the best planning capabilities. We vary the heterogenous worker models from the pool of: Llama-3.1-70B-Instruct, Qwen2.5-32B-Instruct, and GPT-4o-mini. We run a total of 7 experiments with varying combinations of these models for $n=1,2,3$ worker agents. To evaluate the results, we use \ding{182} \textbf{Efficiency} ($=\frac{\#complted\_orders}{\$cost}$) to measure the effectiveness of the task allocation strategies in terms of the number of orders completed relative to the cost incurred.

\begin{wrapfigure}{r}{0.5\linewidth}   
  \centering
  \vspace{-38pt} 
  \includegraphics[width=\linewidth]{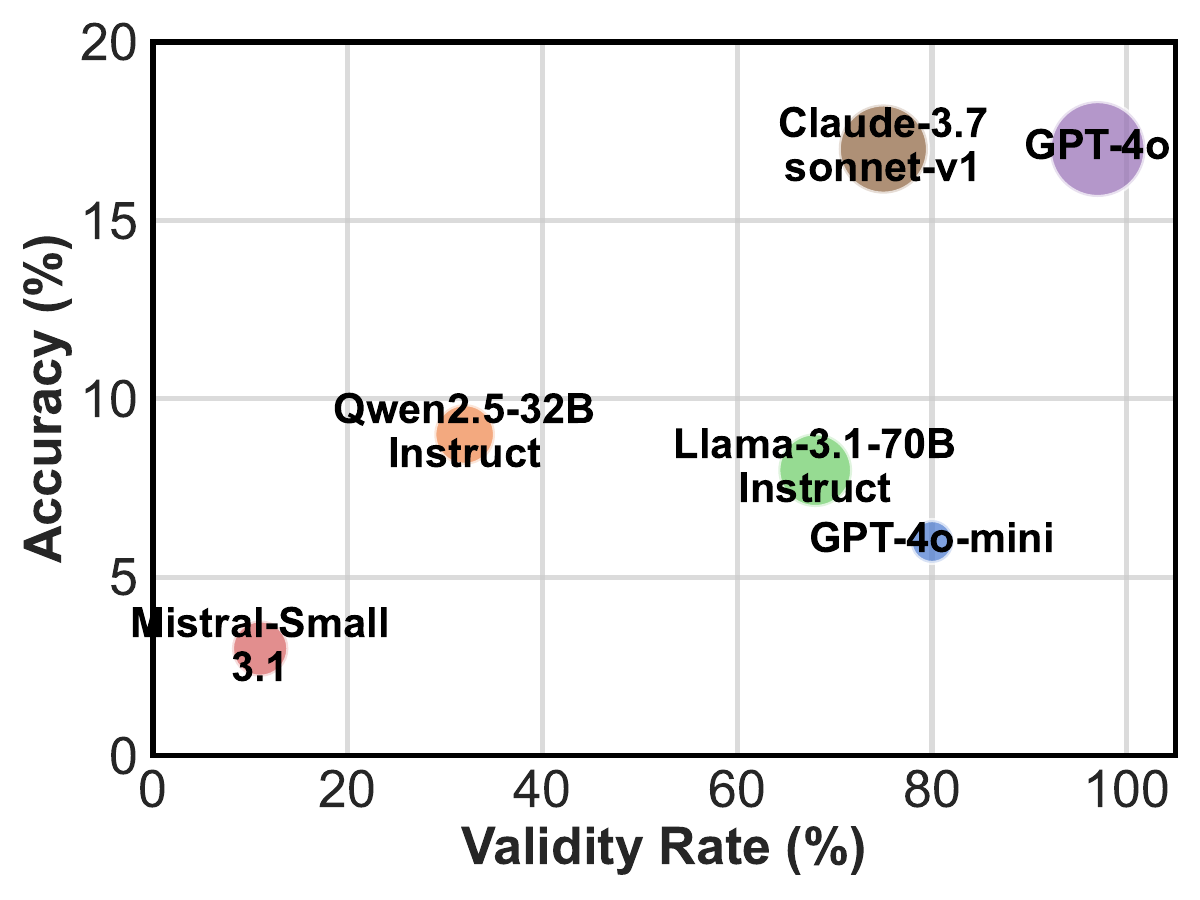}
  \caption{Performance of LLM orchestrators on assignment problems. The circle size scales with the model parameter count. }
  \vspace{-25pt} 
  \label{fig:exp1_result}
  \end{wrapfigure}

\section{Results}
\paragraph{LLMs address the assignment problem, with performance scaling alongside model size and cost.}

As shown in the Figure~\ref{fig:exp1_result}, larger language models generally achieve higher validity and accuracy in resource allocation tasks, indicating that orchestration performance improves with model capacity. However, these gains come at a significant computational and monetary cost: the most capable models are also the most expensive to deploy. Consequently, while an advanced LLM offers superior orchestration abilities, thus raising the need for an alternative more efficient solution for the resource allocation in multi-agent systems.

\begin{figure}[]
    \centering
    \begin{minipage}[b]{0.3\textwidth}
        \centering
        \includegraphics[width=\textwidth]{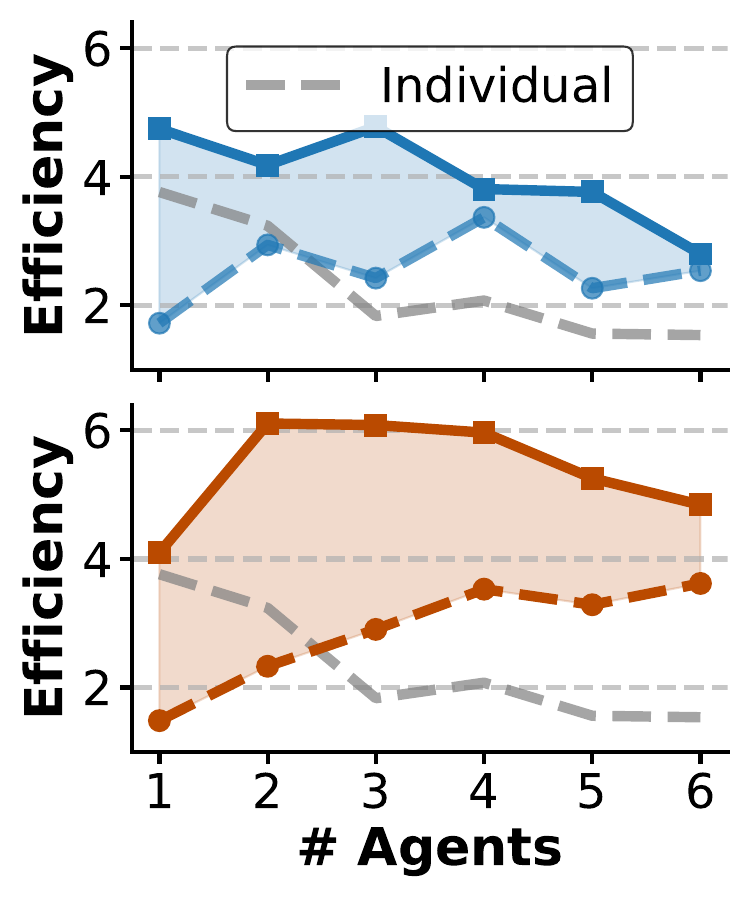}
        \caption{\centering Worker Agent: Llama 70B-Instruct}
        \label{fig:exp2_efficiency_llama70b}
    \end{minipage}
    \hfill
    \begin{minipage}[b]{0.3\textwidth}
        \centering
        \includegraphics[width=\textwidth]{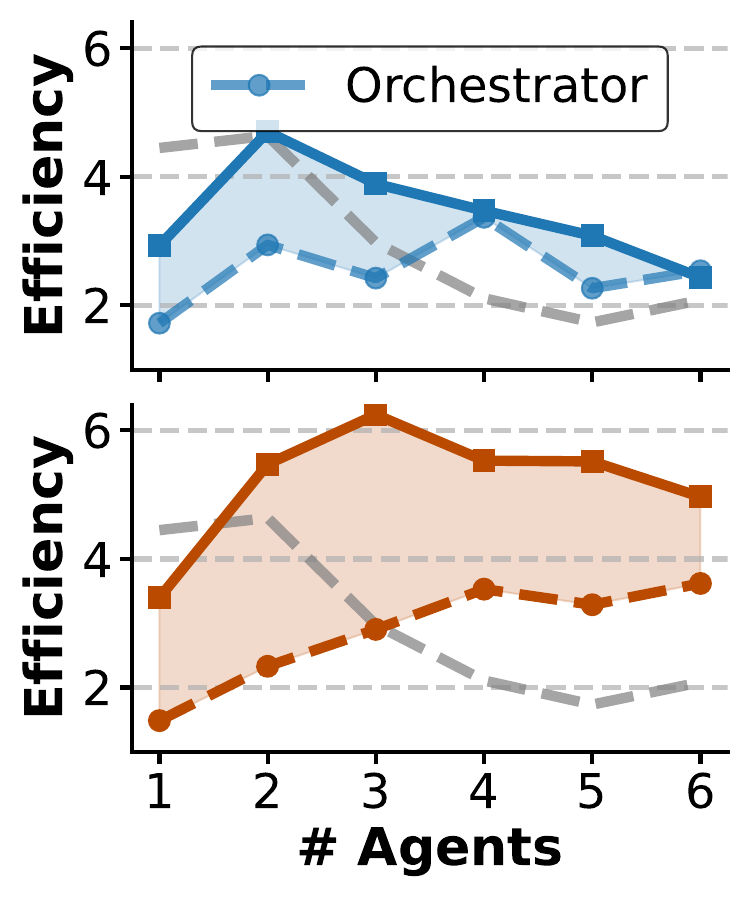}
        \caption{\centering Worker Agent: Qwen-32B-Instruct}
        \label{fig:exp2_efficiency_qwen32b}
    \end{minipage}
    \hfill
    \begin{minipage}[b]{0.3\textwidth}
        \centering
        \includegraphics[width=\textwidth]{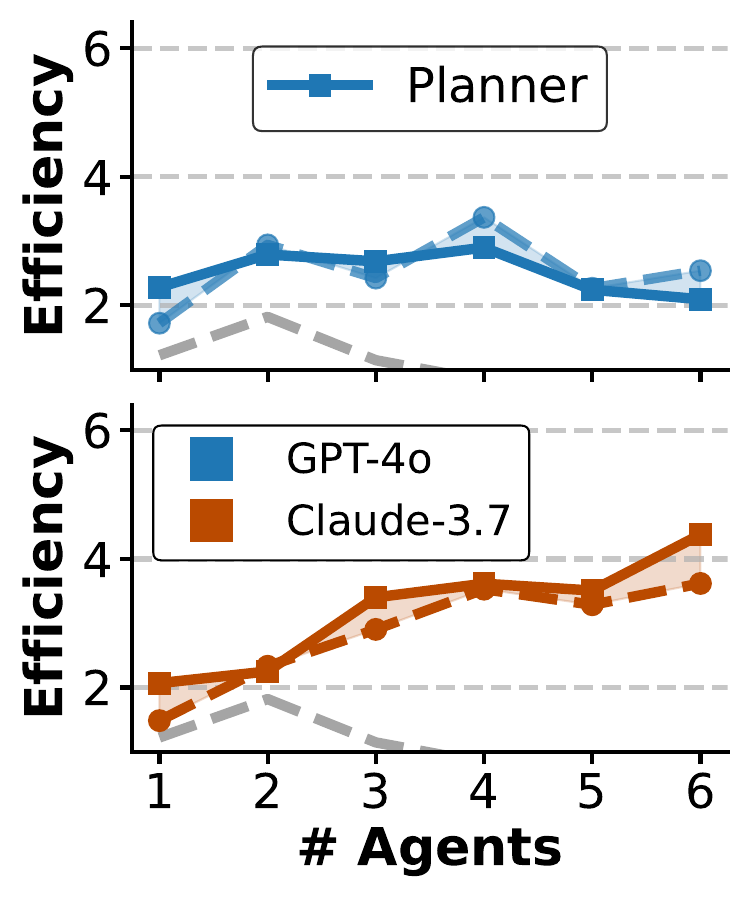}
        \caption{\centering Worker Agent: GPT-4o-mini}
        \label{fig:exp2_efficiency_gpt4o_mini}
    \end{minipage}
    \caption{Efficiency for Planner, Orchestrator and Individual methods. Top (blue) plots use GPT-4o as orchestrator/planner. Bottom plots (brown) use Clause-3.7 as orchestrator/planner. Worker agent models indicated in captions. The results demonstrate the planner method is more efficient than their centralized or decentralized counterparts.} 
    \label{fig:experiment2_efficiency}
\end{figure}

\paragraph{The Planner method achieves better efficiency.} The results are presented in Table \ref{tab:experiment2_results}, where we see that central planning completes the highest number of orders, given that these are the strongest models. These results are expected. However, we want to measure the efficiency of the MAS, specifically what is more efficient given the cost and the number of orders. In Figure \ref{fig:experiment2_efficiency}, we plot the efficiency results for all models. This shows how the Planner method becomes the most cost-efficient for the results obtained.

\begin{figure}[]
    \centering
    \begin{subfigure}[b]{0.38\textwidth}
        \includegraphics[width=\textwidth]{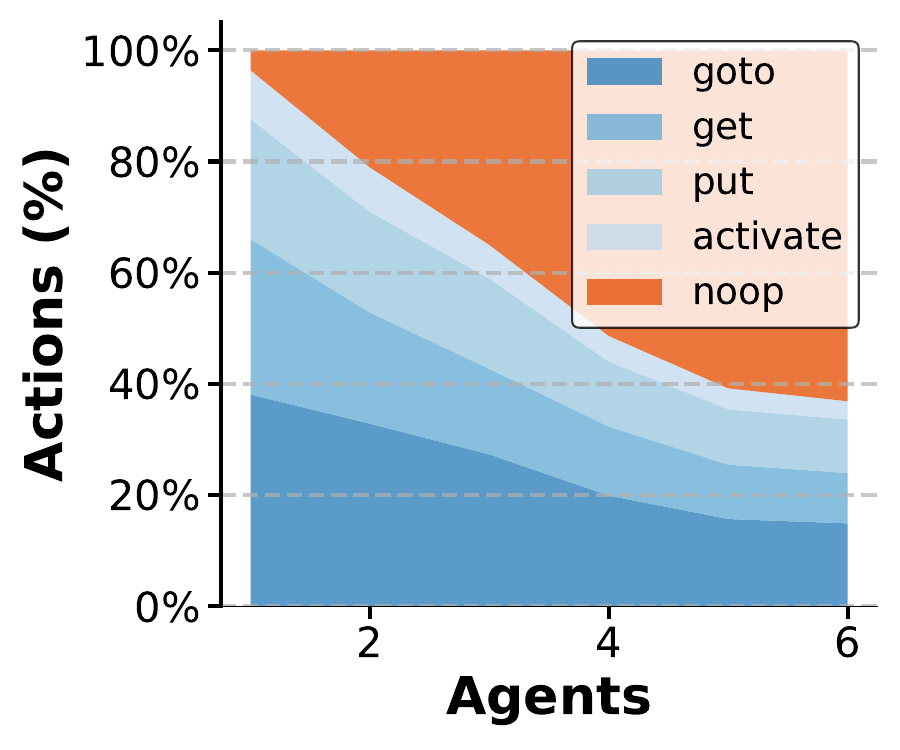}
        \caption{Orchestrator}
        \label{fig:actions_stacked_orchestrator}
    \end{subfigure}
    \hspace{.5cm}
    \begin{subfigure}[b]{0.38\textwidth}
        \includegraphics[width=\textwidth]{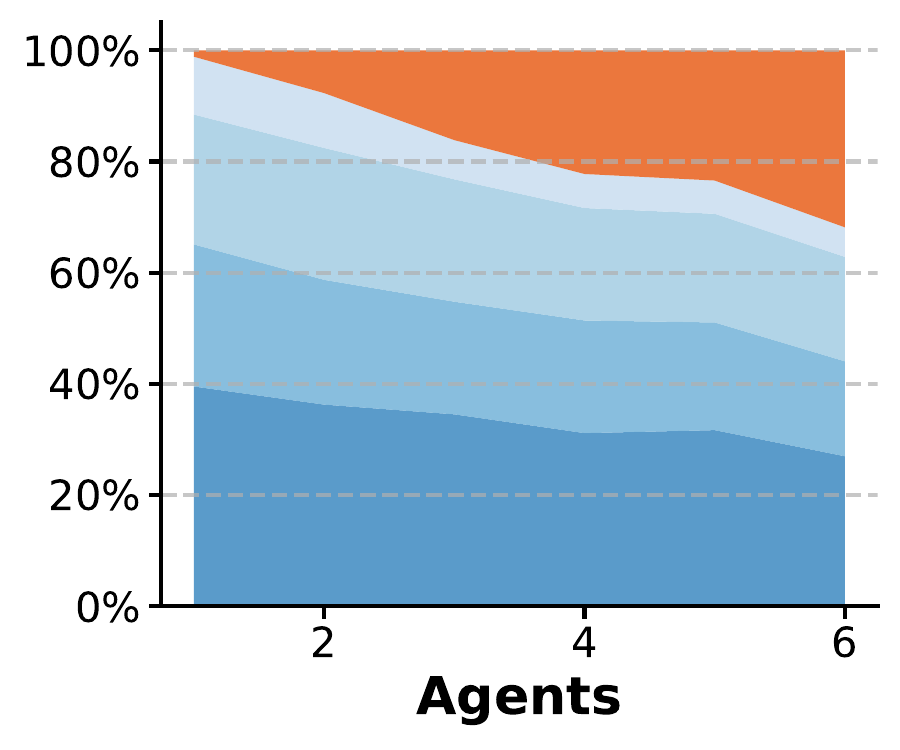}
        \caption{Planner}
        \label{fig:actions_stacked_planner}
    \end{subfigure}
    \caption{Percentage of actions executed by action type in Experiment 2 (\textit{Orchestrator = GPT-4o; Planner = GPT4o+Llama70B-Instruct}). The results indicate that planners maintain a higher percentage of active (non-idle) actions compared to a centralized orchestrator.}
    \label{fig:ex2_actions_stacked}
    \vspace{-14pt}
\end{figure}

Furthermore, in Figure \ref{fig:ex2_actions_stacked}, we see how the Planner method achieves better utilization of the agents, with a lower percentage of idle actions generated. As the agents are given more independence to generate their actions, they tend to produce fewer idle steps than when centrally organized.

\begin{table}[]
    \centering
    \small
    \begin{tabular}{cccccccc|cccccc}
    & & \multicolumn{6}{c|}{\textbf{Completed Orders}} & \multicolumn{6}{c}{\textbf{Cost (\$) }} \\
    \midrule
     \multicolumn{2}{c}{\textbf{Models}} &  \multicolumn{6}{c|}{\textbf{\#Agents}} & \multicolumn{6}{c}{\textbf{\#Agents}} \\
    \cmidrule{3-14}
    \textbf{Plan./Orch.} & \textbf{Worker} & \textbf{1} & \textbf{2} & \textbf{3} & \textbf{4} & \textbf{5} & \textbf{6} & \textbf{1} & \textbf{2} & \textbf{3} & \textbf{4} & \textbf{5} & \textbf{6} \\
    \midrule
    \multicolumn{14}{c}{\textit{Individual}} \\
    \midrule
    \ding{55} & GPT-4o-mini & 1 & 3 & 3 & 3 & 4 & 5 & 0.8 & 1.6 & 2.6 & 3.6 & 4.6 & 5.7 \\
    \ding{55} & Llama-70B & 14 & 25 & 22 & 34 & 33 & 40 & 3.7 & 7.7 & 12.0 & 16.4 & 21.1 & 26.0 \\
    \ding{55} & Qwen-32B & 9 & 20 & 20 & 19 & 20 & 26 & 2.0 & 4.3 & 6.7 & 9.0 & 11.5 & 12.0 \\
    \midrule
    \multicolumn{14}{c}{\textit{Orchestrator}} \\
    \midrule
    GPT-4o & \ding{55} & 20 & 37 & 33 & 48 & 34 & 40 & 11.6 & 12.6 & 13.6 & 14.2 & 15.0 & 15.8 \\
    Claude 3.7 & \ding{55} & 26 & 49 & 66 & 85 & 85 & 98 & 17.5 & 21.0 & 22.7 & 24.1 & 25.9 & 27.1 \\
    \midrule
    \multicolumn{14}{c}{\textit{Planner}} \\
    \midrule
    \multirow{3}{*}{GPT-4o} & GPT-4o-mini & 9	& 11 & 12 & 13 & 12 & 11 & 2.3 & 2.8 & 2.7 & 2.6 & 2.2 & 2.1\\
     & Llama70B & 21 & 27 & 41 & 40 & 48 & 42 & 4.4 & 6.5 & 8.6 & 10.5 & 12.8 & 15.0 \\
     & Qwen32B & 11 & 22 & 22 & 24 & 24 & 22 & 3.8 & 4.7 & 5.7 & 6.9 & 7.8 & 9.0 \\
    \cmidrule{2-14}
    \multirow{3}{*}{Claude-3.7} & GPT-4o-mini & 3 & 9 & 17 & 20 & 22 & 30 & 1.5 & 4.4 & 5.0 & 5.5 & 6.3 & 6.8 \\
     & Llama-70B & 21 & 44 & 57 & 68 & 72 & 77 & 5.1 & 7.2 & 9.4 & 11.4 & 13.7 & 15.9 \\
     & Qwen32B & 16 & 30 & 42 & 43 & 50 & 50 & 4.7 & 5.5 & 6.7 & 7.8 & 9.1 & 10.1 \\
    \bottomrule
    \end{tabular}
    \caption{Results from Experiment 2 (Section \ref{section:experiment2}): Concurrent Allocation in CuisineWorld the number of completed orders and associated costs for different models and methods}
    \label{tab:experiment2_results}
\end{table}

\begin{wrapfigure}{r}{0.55\linewidth}
\centering
\includegraphics[width=\linewidth]{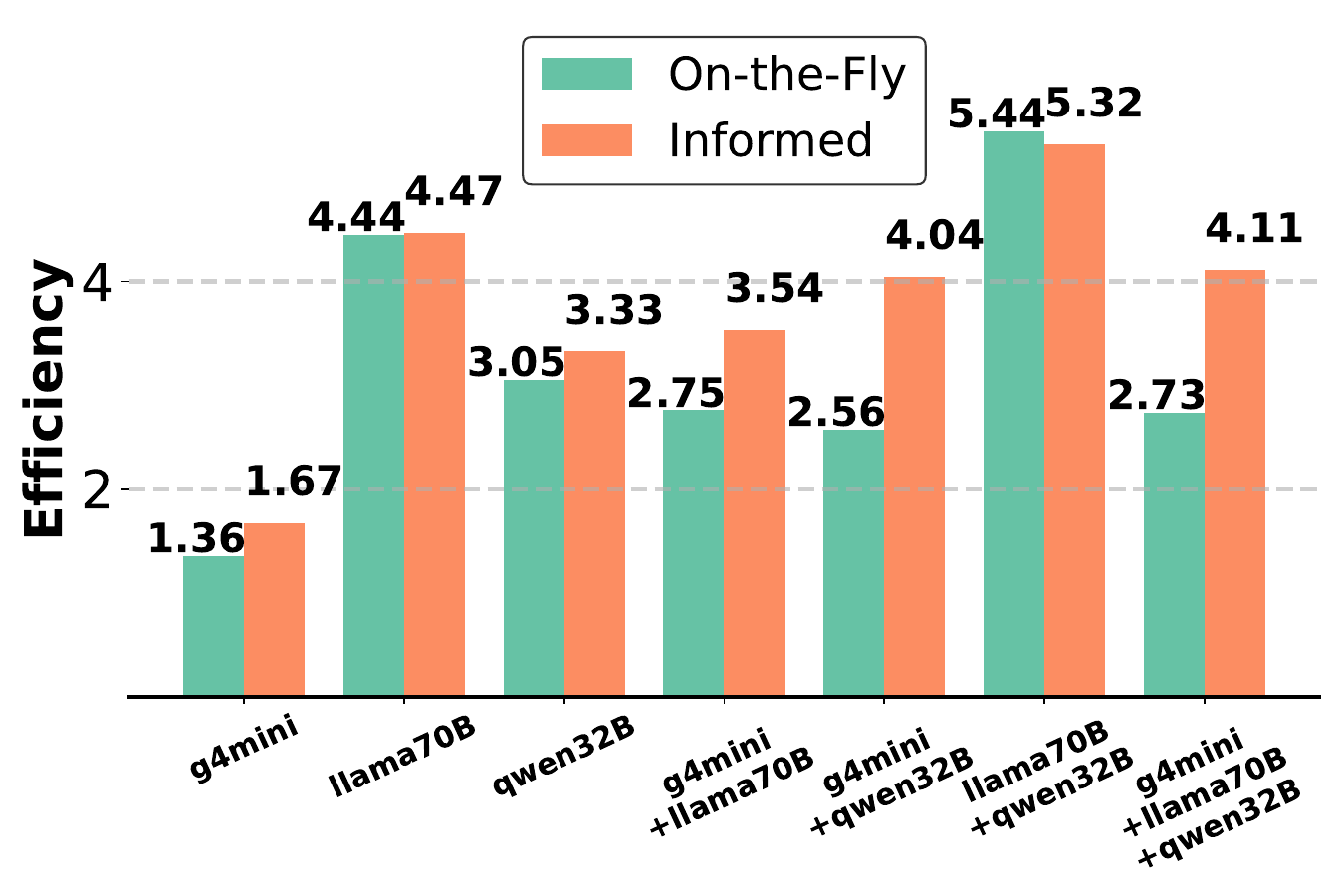}
\caption{Efficiency comparison of On-the-fly Allocation and Informed Allocation for Capability-Aware Task Allocation with LLMs as Planners.}
\label{fig:exp3_efficiency}
\vspace{-10pt}
\end{wrapfigure}

\paragraph{LLMs struggle to infer worker capabilities but improve when given explicit hints.}
Figure \ref{fig:exp3_efficiency} compares On-the-fly Allocation to Informed Allocation. In On-the-fly allocation, Without any prior indication of each worker LLM’s strengths, planners struggle to identify the best allocations dynamically, especially when worker performance varies widely. However, simply providing subtle cues about worker capabilities (action success rate from On-the-fly allocation performance) in Informed Allocation leads to a marked increase in overall efficiency—particularly when planners must work with suboptimal models. These cues reduce non-productive actions, helping LLM-based planners better match tasks to models when aware of worker differences.

\paragraph{LLMs are highly sensitive to worker capabilities.}

We find the performance of a multi-agent system is significantly influenced by the specific capabilities of each worker LLM and how these capabilities are combined. In a homogeneous setting, Llama-70B-Instruct and Qwen32B consistently outperform GPT-4o-mini when used as worker models. Mixing models of varying strengths generally reduces average efficiency. However, including at least one stronger model in a heterogeneous team improves efficiency. For instance, combining Qwen with GPT-4o-mini yields an efficiency of 4.04, compared to 2.79 for two GPT-4o-mini models from experiment 2. In the heterogenous case, we observe that a smaller team of more capable LLMs can outperform a larger team with uneven skills, as seen with the Llama–Qwen pairing outperforming the combination of all three.

\section{Conclusions}

This work explores the capabilities of LLMs to optimize task allocations in multi-agents systems. Experiments show that LLMs can function as orchestrators, with relative performance to that of established algorithms like the Hungarian Algorithm. However, using a planner method instead of an orchestrator improves efficiency in handling concurrent actions. These findings suggest that leveraging LLMs can create more efficient and cost-effective multi-agent frameworks, dynamically allocating resources based on real-time needs, enhancing performance and cost-effectiveness.




\section*{Ethics Statement}

This paper aims to automate processes using AI systems, focusing on developing efficient and intelligent multi-agent frameworks for applications like research assistance and task management. We acknowledge the societal implications of our work and emphasize the need for further analysis to understand LLM outputs, especially in real-world applications. We advocate for transparent algorithmic processes that allow individuals to comprehend AI-driven decisions. Our experiments are conducted in a controlled gaming environment to ensure safety and prevent unintended impacts on other systems.

\bibliography{colm2025_conference}
\bibliographystyle{colm2025_conference}

\newpage
\appendix
\section{CuisineWorld}
\label{app:cuisineworld}

\begin{figure}[!h]

    \centering
    \begin{minipage}{0.45\textwidth}
    \centering
    \begin{tabular}{ccc}
        \toprule
        \textbf{Type} & \textbf{Arguments} & \textbf{Description} \\ 
        \midrule
        \multirow{2}{*}{\texttt{goto}}   &   \texttt{agent}  &   Move \texttt{agent} \\
        & \texttt{location} &   to \texttt{location} \\
        \midrule
        \multirow[c]{3}{*}{\texttt{get}} & \texttt{agent} &  \texttt{agent} obtain\\ 
        & location & \multirow[c]{2}{*}{\texttt{item} from \texttt{location}} \\ 
        & (\texttt{item}) & \\  
        \midrule
        \multirow{2}{*}{\texttt{put}}   &   \texttt{agent}  & \texttt{agent} put everything \\
        & \texttt{location} &    it holds to \texttt{location} \\
        \midrule
         \multirow{2}{*}{\texttt{activate}}   &   \texttt{agent}  & \texttt{agent} turn on \\
        & \texttt{location} &  \texttt{location} \\
        \midrule
        \multirow{2}{*}{\texttt{noop}}   &   \texttt{agent}  & not dispatching \\
        & \texttt{location} &  \texttt{agent} \\
        \bottomrule
    \end{tabular}
    \captionof{table}{Action space in CuisineWorld.}
    \label{tab:cuisineworld_action_space}
    \end{minipage}
    \hfill
    \begin{minipage}{0.4\textwidth}
    \centering
        \includegraphics[width=\linewidth]{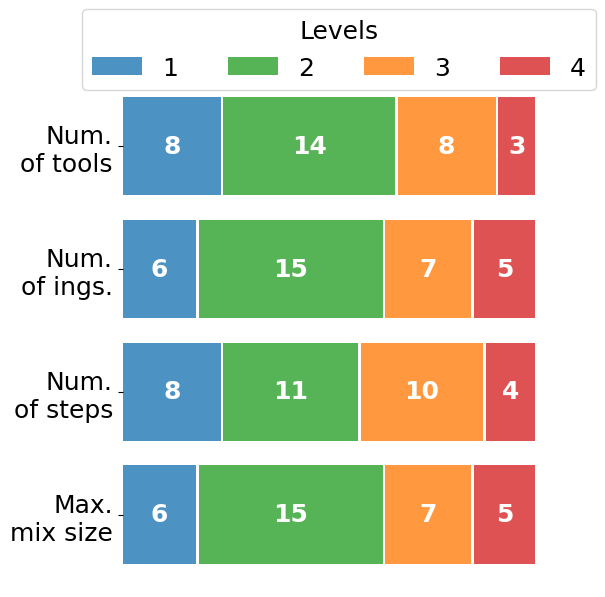} 
        \caption{Dish distribution over the number of tools and ingredients (ings.) involved, cooking steps, and maximum mixture size as in the recipe.}
        \label{fig:cuisineworld_dish_distribution}
    \end{minipage}
\end{figure}

CuisineWorld \citep{mindagent} is a benchmark designed to evaluate the planning and coordination capabilities of multi-agent systems. It simulates a virtual kitchen environment where multiple agents need to collaborate to complete various cooking tasks. actions that can be taken by the LLMs are described in \ref{tab:cuisineworld_action_space}. Recipes provide a step-by-step guide for preparing different dishes. They list the required ingredients for each part of the process, the tools you'll need, and what the final dish should look like once it's cooked. The recipes are grouped into 13 levels, with different range of difficulty depending on the number of cooking tools involved, the number of ingredients, and the number of steps required to complete the dish. Figure \ref{fig:cuisineworld_dish_distribution} shows the distribution of these factors over these groups.

Data provided from the environment is given to the agents in text form, as shown in Table \ref{tab:cuisineworld_gamestate}.

\begin{table}[!h]
    \small
    \begin{tabular}{ll}
    \toprule
    \textbf{Game Configuration} & \\
    \midrule
    Current Game Level & \texttt{level\_1} \\
    Current Dishes & \\
    \quad Name & \texttt{salmonMeatcake} \\
    \quad Lifetime & 10 \\
    Current Game Step & 10 \\
    Maximum Game Steps & 60 \\
    \midrule
    \textbf{Agent State} & \\
    \midrule
    \texttt{at(agent0, servingtable0)} & \\
    \texttt{hold(agent0, None)} & \\
    \texttt{at(agent1, servingtable0)} & \\
    \texttt{hold(agent1, None)} & \\
    \midrule
    \textbf{Kitchen State} & \\
    \midrule
    \texttt{inside(storage0, None)} & \\
    \texttt{inside(servingtable0, None)} & \\
    \texttt{inside(blender0, None)} & \\
    \texttt{inside(blender1, None)} & \\
    \midrule
    \textbf{Accomplished Tasks} & \\
    \midrule
    \texttt{salmonMeatcake} & \\
    \bottomrule
\end{tabular}
    \caption{Example of Game State returned from the environment and provided to the agents.}
    \label{tab:cuisineworld_gamestate}
\end{table}


\section{Model Prices}

In Table \ref{tab:models_prices}, we present the models' prices selected at current rates from their official APIs. OpenAI API price website: \href{https://openai.com/api/pricing/}{openai.com/api/pricing/}, Anthropic API pricing: \href{https://www.anthropic.com/pricing}{anthropic.com/pricing}. For the case of Open-weights, we take the values from Llama-API provider: \href{https://www.llama-api.com/pricing}{llama-api.com/pricing}. All websites have been checked at date 25 March 2025.

\begin{table}[!h]
    \centering
    \begin{tabular}{lcccc}
    \toprule
    \multirow{2}{*}{\textbf{Model}} & \multirow{2}{*}{\textbf{Company}} & \textbf{Open} & \textbf{Input Cost} & \textbf{Output Cost} \\
    & & \textbf{Weights} & \textbf{(\$/M tokens)} & \textbf{(\$/M tokens)} \\
    \midrule
    claude-3.7 & Anthropic & \ding{55} & 3.00 & 15.00 \\
    \midrule
    gpt-4o-v2 & OpenAI & \ding{55} & 2.50 & 10.00 \\
    \midrule
    gpt-4o-mini & OpenAI & \ding{55} & 0.15 & 0.60 \\
    \midrule
    Llama-3.1-70B-Instruct & Meta & \ding{51} & 0.80 & 2.80 \\
    \midrule
    Qwen2.5-32B-Instruct & Alibaba & \ding{51} & 0.40 & 1.40 \\
    \bottomrule
    \end{tabular}
    \caption{Model Prices from providers.}
    \label{tab:models_prices}
\end{table}

\end{document}